# The unified electronic nature of nanomaterial surface science

Guolei Xiang

Department of Energy Chemistry, College of Chemistry, Beijing University of Chemical Technology, Beijing, China, 100029

Email: xianggl@buct.edu.cn; xgl8512@gmail.com.

**Abstract:**

The properties of inorganic nanomaterials in adsorption, catalysis, and photoluminescence are commonly affected or dominated by particle size, surface ligand, and ligand coverage; however, it has been long remaining challenging to generally understand underlying physical and chemical principles with a unified model. In this review the electronic-level principle that can unify the structure-property relationships concerning nanomaterial surface science is systematically illustrated, based on a chemisorption model through competitive orbital redistribution from bulk energy bands into surface chemisorption bonds. The physical nature of enhanced surface reactivity by size reduction lies in weakening lattice confinement on surface atomic orbitals and amplifying the effects of other structural factors such as defects. Nanoscale cooperative chemisorption (NCC) model reveals the general physical principles driving size- and coverage-dependent ligand-nanomaterial interactions owing to orbital competitions between bulk energy bands and surface adsorption bonds. NCC theory can interpret the impacts and trends of ligand-induced surface effects on electronic states, bonding strength, and the emission energy and quantum yield of both surface fluorescence like ligand-capped Au nanoclusters and bulk intraband fluorescence like CdSe quantum dots. Competitive orbital redistribution provides a new electronic perspective to unify the physical principles driving nanomaterial surface science.

## Introduction

Nanomaterials all geometrically display ultrasmall sizes and large surface-to-volume ratios (*S/V*)[1]. This feature enables surface chemical states to highly affect or even dominate their structures and properties and leads to various ligand-induced surface effects (LISEs)[2,3]. Meanwhile, the surface science of nanomaterials also widely underlies adsorption, heterogeneous catalysis, photocatalysis, electrochemistry, photovoltaic device, chemical sensing, photoluminescence, as well as the synthesis, assembly, and functionalization of nanocrystals[4-8]. As a result, nanomaterial surface science in the interface where physics meets chemistry, and plays central roles in nanoscience and nanotechnology[2,5,9]. Revealing the fundamental physical and chemical principles driving nanomaterial surface science is then crucial to understand the unusual nanoscale phenomena and interactions[1,10]. Although several models have been proposed to explain the origins of nano effects that basically result from nanoscale size and inversely proportional *S/V* (Figure 1A), it has been long remaining challenging to unify the origin, feature, and interaction trend of nano effects on nanomaterial surface science[1,2,11].

The scope and feature of nanomaterial surface science markedly differ from that of single crystals[2]. More structure-property relationships are involved owing to the diverse structural variables such as size, *S/V*, facet, shape, defect, strain, interface, capping atom and coverage of ligands (Figure 1B)[12,13]. Among these factors, size effect and LISE are the most typical nano effects illustrating the features of nanomaterial surface science[1,2,13]. Despite the differences in chemical compositions and shapes of nanomaterials, interaction strength and impacts of ligand-nanomaterial interactions can be enhanced by decreasing particle size and increasing ligand coverages. This is the general interaction trend of nanomaterial surface science.

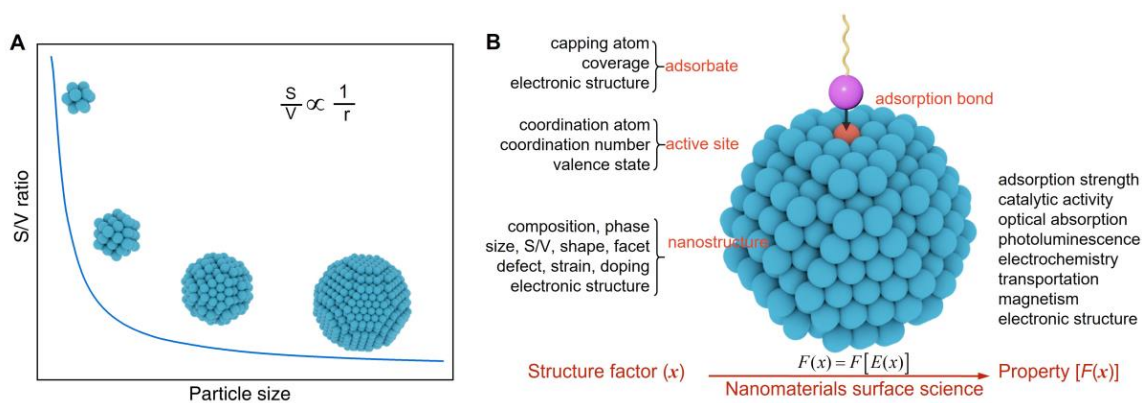

**Figure 1.** The scope and features of nanomaterial surface science. (**A**) The basic geometric feature of inverse scaling of *S/V* with size of nanomaterials. (**B**) The scope and structure-property relationships in nanomaterial surface science.

## The general feature and trend of nanomaterial surface science

Size effects in nanomaterial surface science display as size-dependent performances in adsorption, catalysis, sintering, phase transformation, and optical absorption and emission (Figure 1B)[1,6,14-16]. Size reduction can generally increase surface activities[1,10], which is typically illustrated by Au catalysis and the decreased melting point of metal nanoparticles (NPs) from more than 1000 °C down to about 200 °C[16-19]. Commonly understanding the physical nature of such size effects is a central fundamental issue of nanoscience, but has not been unified due to challenges in theory and experimental probing[1,10,11,19,20].

Surface effects on nanomaterials mainly display as the modifications of physical and chemical properties by surface ligands[4,13,21]. LISEs include three types according to the contributions to properties: ligand-dominant property like solubility[22,23], nanomaterial-dominant property like superparamagnetism of $Fe_3O_4$ NPs and photoluminescence of quantum dots (QDs)[24,25], and chemisorption-dominant property like the photoluminescence of ligand-capped metallic nanoclusters (NCs)[26]. Most properties of nanomaterials can be modified by ligands through chemisorption, and the typical variables include particle size, and the capping atom and coverage ($\theta$) of ligands[27,28]. In general, the results of LISEs can be enhanced by increasing $S/V$ and ligand coverage. This interaction trend can be typically illustrated by the photoluminescent performances of ligand-capped Au NCs and CdSe QDs[25,26].

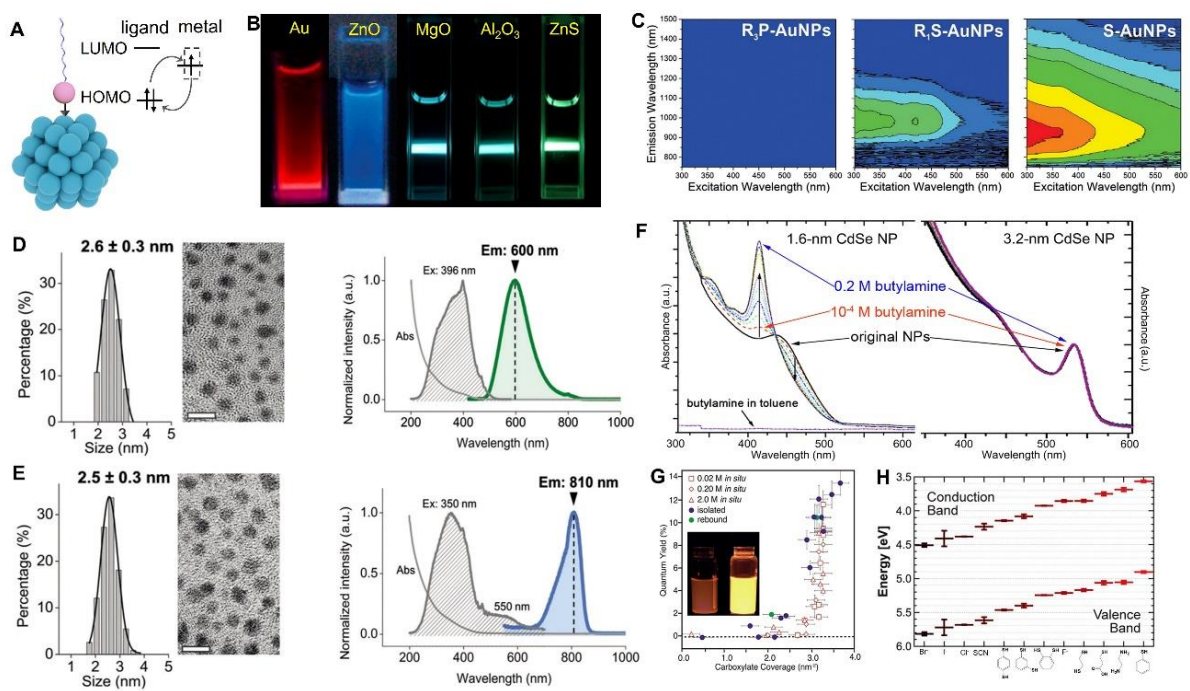

**Figure 2**. Ligand- and coverage-dependent surface effects on the optical properties of nanoclusters (NCs) and quantum dots (QDs). (**A**) Surface optical absorption and emission of NCs resulting from electron transfer between ligands and metal sites. (**B**) Surface photoluminescence of ligand-capped NPs of Au, ZnO, MgO, $Al_2O_3$, and ZnS. (Permission from ). (**C**) Ligand-dependent photoluminescence of Au NCs. (**D-E**) Size-independent but cover-dependent photoluminescence of Au NCs. (**F**) Size effect on the optical absorption of ligand-capped CdSe NPs. (**G**) Coverage-dependent photoluminescent quantum yield of CdSe QDs. (**H**) Ligand effects on the band edges of PdS NPs.

Surface ligation of nanomaterials can generate new electronic states for optical absorption and emission[26,29]. These modes result from the surface bonding states of ligands with nanomaterials through chemisorption[9,30]. Optical absorption usually results from irradiated electron transfer from ligands to metals, while emission results from relaxation of excited electrons from metals to ligands (Figure 2A)[29,31]. Nanomaterials can be metals, oxides, and semiconductors[30,32,33] (Figure 2B), while the capping groups of ligands are usually -SH, -NH$_2$, and –COOH[2,34].

The photoluminescent wavelength and quantum yield (PLQY) of thiol-capped Au NCs can be tuned by changing particle size, and the capping atom and coverage of ligands[35]. (i) Particle size primarily affects photoluminescent wavelength[31,36]. For example, the emission maximum can shift from ultraviolet (UV) to infrared region when increasing amine-capped Au NCs from Au$_5$ to Au$_{31}$, and from 800 to 1300 nm when increasing thiol-capped Au NCs from Au$_{11}$ to Au$_{38}$, Au$_{140}$ and Au$_{201}$[37-39]. The evolution trends indicate that decreasing particle size can increase photoluminescent energy. (ii) Capping atoms can control the on-off of surface fluorescence[40]. For example, Au NCs capped by phosphine (P) ligands show no fluorescence in near infrared (NIR) region when excited by irradiation from 300 to 600 nm; however, photoluminescence can be switched on through exchanging P-ligands with thiols and sulfide, which leads to emission between 800 to 1300 nm (Figure 2C)[41]. This change shows that the type of ligation bonds can critically determine the on-off of surface fluorescence. (iii) Ligand coverage can tune both emission wavelength and PLQY of glutathione-capped (GSH) Au NPs of ~2.5 nm (Figure 2D, 2E)[42]. Low GSH coverage leads to an emission maximum at 810 nm with a 1.1% PLQY (Figure 2D), while high GSH coverage leads to a single emission at 600 nm with a 5.3% PLQY (Figure 2E). Such size-independent emissions indicate that ligand coverage can critically control the emission energy and PLQY of thiol-capped Au NCs. These examples conclude a trend that decreasing particle size or increasing ligand coverage can enhance both the emission energy and quantum yield of surface fluorescence of Au NCs.

Particle size, and ligand's capping atom and coverage can also affect the bulk optical absorption and emission of semiconductor QDs resulting from electron transfers between valence bands (VB) and conduction bands (CB)[43-46]. (i) The colloidal solution of 1.6-nm CdSe NPs capped by trioctylphosphine (TOP) shows an absorption peak at 445 nm, but the addition of *n*-butylamine leads to a narrower absorption maximum at 414 nm. In contrast, *n*-butylamine cannot change the absorption spectra of 3.2-nm CdSe NPs (Figure 2F)[47,48]. These results show that particle size and capping atoms of ligands can affect the surface optical absorption of QDs and the features of surface electronic states. (ii) Ligand coverage can also influence the fluorescent efficiency of QDs[49]. For example, the PLQY of CdSe QDs capped with oleic acids (OA) tremendously drops when surface OA density decreases from 6 nm$^{-2}$ to 3 nm$^{-2}$ (Figure 2G)[2]. The PLQY can be further quenched by decreasing OA coverage below 2 nm$^{-2}$, and can also be decreased by the loss of surface ligands during purification or dilution of CdSe NPs solutions[50]. (iii) The bulk photoluminescence of NaGdF$_4$:Yb/Tm and perovskite CsPbBr$_3$ NPs can also be enhanced through surface ligation. Surface ligation of NaGdF$_4$:Yb/Tm NPs with bidentate picolinic acid molecules can enhance four-photon upconversion by11,000-fold[51]. The external quantum efficiency of quantum dot light-emitting diodes (QLEDs) can be improved for 50-fold for solution-processed all-Inorganic perovskite CsPbBr$_3$ NPs by controlling surface ligand densities[52]. As a result, these absorption and emission results show that surface ligands can also enhance the bulk photoluminescent properties.

Surface ligands can also modify the energy levels of valence band maximum (VBM) and conduction band minimum (CBM) of QDs, and further change the band gaps[53-56]. For example, the energy levels of PbS QDs can be shifted by up to 0.9 eV after ligand modification (Figure 2H)[55]. Surface ligation is also an important approach to control the electronic properties of QDs and to optimize their performances in optoelectronic devices, in complement to quantum size effects on modifying their band gaps and electronic structures[52,57-61]. Such shifts were thought to result from the contributions of both the QD-ligand interface dipoles and intrinsic dipole moment of ligand molecules[29,62].

## Common issues and challenges in nanomaterial surface science

Regardless of the differences in chemical compositions of nanomaterials and ligands, numerous experimental explorations have clarified that decreasing particle size and increasing ligand coverage can enhance ligand-nanomaterial interactions. Underlying these progresses in phenomena, many fundamental questions have long remained open as to the general physical principles driving nanomaterial surface science[10,11]. Although size reduction down to the bottom region of nanoscale (typically < 5 nm) is definitely the direct cause of all nano effects, as a geometric-level parameter it is not the intrinsic nature at all. To unify the interpretations on the features and trends of nanomaterial surface interactions, the physical nature of the following issues needs to be addressed in depth. (i) How adsorbates interact with nanomaterials and mutually change their properties. (ii) How size reduction commonly modifies the intrinsic electronic structures and enhances surface energy and surface activity of nanomaterials. (iii) The roles particle size, *S/V*, and ligand coverage in adsorbate-nanomaterial interactions. In addition, some experimental phenomena also need to be intrinsically explained. For example, why the surfaces of nanocrystals prepared through wet-chemistry methods (such as QDs and Au NCs) tend to be saturated by ligands; why and how high ligand coverages can enhance both the surface fluorescence of Au NCs and bulk fluorescence of QDs[63-65]. A more ambitious question might be if these issues can be unified in a theory[9,11].

The structure-property relationships in nanomaterial surface science have been discussed at the geometric, atomic, and electronic levels[20,66,67]. (i) Geometric-level structural parameters include particle size, shape, and *S/V*. Size reduction is the direct cause of all nano effects, shape can vary the exposed facets, and increased *S/V* increases the probability for adsorbates to interact with nanomaterials[10]. However, they cannot uncover the intrinsic mechanisms why nanomaterials generally display high surface activities, nor interpret the origins of coverage-dependent surface effects. (ii) Atomic-level parameters include facet, defect, dopant, coordination atom and number, and are widely used to explain various nano effects by chemists[68]. It is also thought that high surface activity results from increased low coordinated sites at the surfaces, edges, kinks, and corners of nanocrystals. Changes of these atomic structures can practically modify local chemical environments of surface sites, and further tune the adsorption and catalytic performances of nanomaterials[19,20]. Although these atomic-level parameters can account for enhanced surface reactivity, they cannot explain the features and trends of size effects and surface effects. (iii) In contrast, electronic-level insights are intrinsic, as all physical and chemical properties of nanomaterials are driven by electronic structures[69]. Therefore, revealing the electronic features of nanomaterials and the intrinsic correlation to surface chemical interactions is the ultimate way to develop the general model and theory[70-73]. Two electronic concepts, quantum size effect and oxidation state, have been widely used to discuss the electronic features of nanomaterials and the mechanisms of nanomaterial surface science[74].

Quantum size effect refers to the discretization of energy levels in energy bands when particle sizes shrink down to the nanometer scale[75,76]. This effect is the reversed evolution of electronic structures from discrete levels of atoms and molecules to continuous energy bands of crystals as shown in Figure 3A. R. Kubo first explained size-induced metal-to-insulator transition of metallic clusters with such size-dependent discretization of band levels[77]. According to the Kubo effect, nanocrystals are regarded as the transition states between molecules and solids[75]. Such quantized band states undoubtedly depict the key electronic features of nanomaterials. However, the principles how such quantized energy bands further correlate to surface activity, size effect, and surface effect remain unclear.

Oxidation state is the most commonly used electronic concept to explain surface chemical interactions by chemists[70]. For example, it was suggested that the catalytic activity of Pd clusters in CO oxidation strongly correlated with cluster size and oxidation states[69]. However, as a state function describing the chemical and electronic states of specific atoms, oxidation state is not intrinsic to interpret surface reactivity, nor to reveal the nature and origin of size effect and surface effect. On the other hand, d-band center model has been successful in interpreting the catalytic tendencies of metals through calculations based on density functional theory (DFT)[71,78]. However, this method cannot be simply extended to compound materials like oxides and nanomaterials[72,73]. Understanding the electronic-level physical nature how adsorbates interact with nanomaterials has long remained a challenging fundamental issue, particularly the electronic states of ligand-capped nanocrystals[11,26,29,70]. These long-term challenges suggest that, to reveal the general physical and chemical principles driving nanomaterial surface science, new physical model, concepts, and theory need to be more insightfully established[2,9,11].

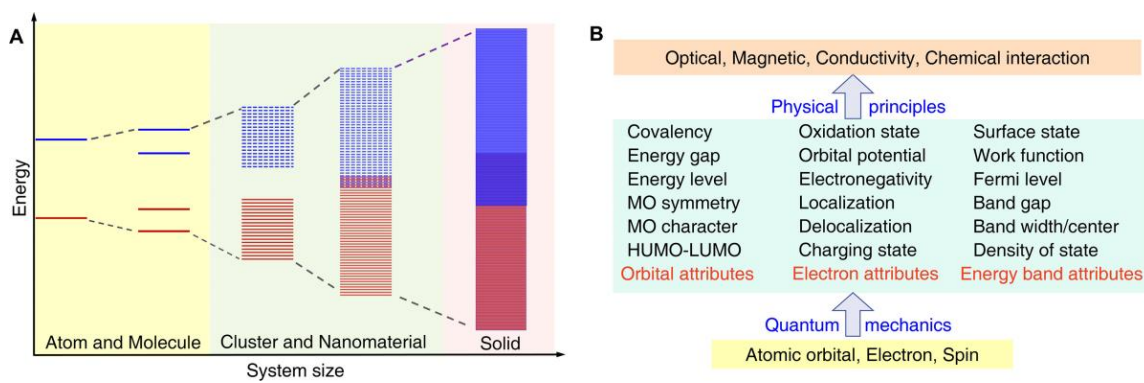

**Figure 3.** Electronic structure attributes. (**A**) Electronic structure evolution from atom to cluster and solid. (**B**) Electronic attributes determining physical and chemical properties.

### Electronic principle of structure-property relationships

Electronic structure intrinsically determines most physical and chemical properties of nanomaterials, while they are also simultaneously influenced by various geometric and atomic structural parameters. Such structure-property relationships can be mathematically expressed as

$$F(x) = F[E(x)] \quad (1),$$

where $F$ denotes physical and chemical properties, $E$ denotes the characteristic electronic attribute (CEA, Figure 3B) that directly determines a specific property, and $x$ denotes structural parameters. Then structural effects (such as size effect) can be expressed as the derivative of $F$,

$$\frac{\partial F}{\partial x} = \frac{\partial F}{\partial E} \cdot \frac{\partial E}{\partial x} \quad (2).$$

The equations illustrate two aspects to rationally reveal structure-property relationships, in which the characteristic electronic attribute $E(x)$ acts as a critical bridging quantity. Clarifying the CEA that directly determines a specific $F$ is the precondition to thoroughly understand the physical nature of all structure-property relationships (Figure 3B)[79]. For example, the gap between energy levels is the CEA that directly determines the wavelengths of optical absorption and emission.

To reveal the electronic principles driving nanomaterial surface science, two concepts need to be clarified. One is the appropriate descriptor denoting the key feature of surface chemical interactions, while the other is the CEA that directly controls surface activity and chemical reactivity. Although adsorption energy is widely used to measure surface activity and reactivity, the value varies with adsorbates and cannot reflect detailed interaction mechanisms as a scalar quantity[72]. Oxidation state is widely used to discuss the electronic mechanisms of surface reactivity, but it is not the CEA directly determining surface reactivity. Therefore, the electronic-level physicochemical picture and key feature of surface chemical interactions need to be clarified[80].

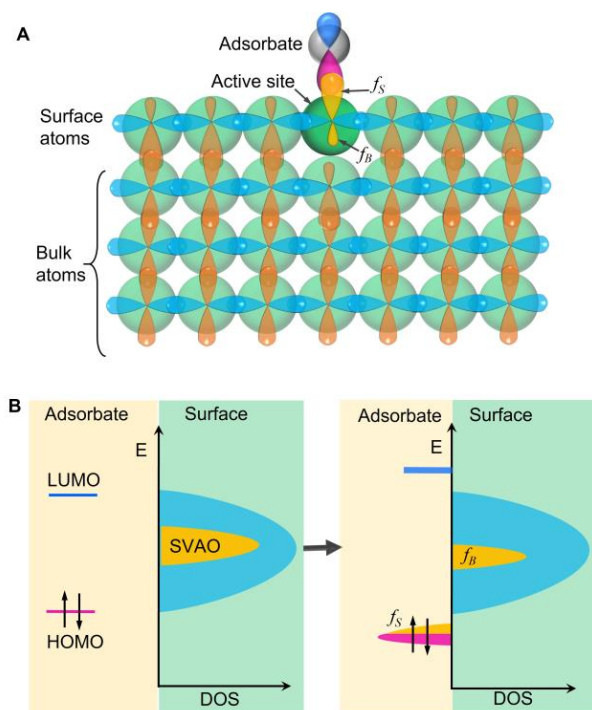

**Figure 4.** Physical model of chemisorption based on competitive orbital redistribution. (**A**) Orbital redistribution of surface valence atomic orbital (SVAO) from extended energy band

states in the bulk phase into localized surface bonds. (**B**) DOS diagram of two-state distributions of SVAO induced by chemisorption.

## A general electronic model of chemisorption

Chemisorption, occurring through forming surface coordination bonds between adsorbates and surface atoms, is the basis of all surface chemical interactions[81-83]. Catalysis involves dynamic chemisorption, while surface ligations are based on static chemisorption. In all chemisorption, the frontier molecular orbitals (FMOs) of adsorbates overlap with the valence atomic orbitals (VAOs) of surface atoms, which yields new surface electronic states (Figure 4A)[79].

The strength of surface coordination bonds is the key feature of chemisorption, and the common character of surface chemical interactions. For given adsorbate and material, the variation of chemisorption strength can be measured by the overlapping degree of their orbitals. In general, VAOs of surface atoms (SVAOs) can distribute into two electronic states: overlapping with VAOs of bulk atoms to extend into energy bands, or overlapping with adsorbate' FMOs to be localized into surface coordination bonds. The distribution fractions (*f*) of a specific SVAO can be denoted as $f_S$ and $f_B$, which measure the contributions of the SVAO to surface bond and energy bands (Figure 4A). For given adsorbate and nanomaterial, $f_S$ is a descriptor of chemisorption strength and surface reactivity, while $f_B$ represents the confinement strength of SVAOs by energy bands and other lattice atoms, and the stability of surface atoms[79]. Then structure-reactivity relationship can be expressed as

$$f_S(x) = f_S[E(x)] \quad (3).$$

The physical nature of $f_S$ and $f_B$ is the square of atomic orbital coefficient in combination states. Electronic orbitals are normalized according to the principles of quantum mechanics, then

$$f_S + f_B = 1 \quad (4).$$

As $0 \leqslant f_S, f_B \leqslant 1$, $f_S$ and $f_B$ are competitive, then $\Delta f_S = -\Delta f_B$. The formulas reveal the physical nature and origin why size reduction can generally lead to increased surface activity but decreased stability. This is because surface activity and stability of nanomaterials originate from the two competitively distributed states of one SVAO.

Chemical interactions generally occur through forming new bonds but breaking old ones. The more fundamental physical process is to redistribute the combination states of atomic orbitals from old bonds to new bonds. For surface chemical interactions, SVAOs are partly redistributed from bulk energy bands into surface chemisorption bonds—this is the central electronic process and direct result of chemisorption[84]. As a result, ligand-capped nanomaterials display two competitive electronic states, extended energy bands in the bulk and localized surface chemisorption bonds. The two electronic states can overlap in energy levels, but are spatially separated (Figure 4B). Competitively redistributing SVAOs from energy bands into chemisorption bonds is the general mechanism how ligands modify the electronic structures and properties of nanomaterials.

Chemisorption can be regarded as a steady state, thus can be studied using the method of state function as in thermodynamics. For steady chemisorption, $f_s$ and $f_B$ of a specific SVAO display a fixed ratio and follow

$$\frac{f_S}{f_B} = \frac{G_S}{G_B} \quad (5).$$

Here $G_S$ and $G_B$ are the CEAs of adsorbates and surface sites that directly determine their capabilities to constrain SVAOs in forming chemisorption bonds. The physical quantity $G$ is defined as orbital potential, and is the electronic attribute measuring the contribution of an atom to a covalent bond locating at any chemical environments (Figure 4B)[79]. In chemisorption bonds, the atoms are ligands' capping atoms and surface atoms acting as active sites. It is noted that $G$ is a state function depending on the surrounding chemical environments of interacting atoms. For surface atoms, $G_B$ is site-dependent and represents the confined strengths of SVAOs by energy bands. For an isolated single atom without other bonding atoms, $G$ becomes electronegativity, a fundamental concept defined by Linus Pauling in 1932. Therefore, orbital potential (G) is the generalized form of electronegativity, and describes the contribution of an atom to a bonding state in any chemical environments.

Equations (4) and (5) derive

$$f_S = \frac{G_S}{G_S + G_B} \quad (6).$$

This equation shows that chemisorption strength depends on the ratio of $G_S/G_B$. Increase of $G_S/G_B$ through increasing $G_S$ or decreasing $G_B$ can enhance chemisorption strength. For given adsorbate and nanomaterial, $G_S$ remains constant, and then $G_B$ totally determines surface reactivity. Therefore, $G_B$ is the CEA directly determining surface reactivity. Accordingly, structure-reactivity relationship can be converted into

$$f_S(x) = f_S[G_B(x)] \quad (7).$$

As a result, the structure-reactivity relationship ($\frac{\partial f_S}{\partial x}$) is transformed into revealing $\frac{\partial f_S}{\partial G_B}$ and $\frac{\partial G_B}{\partial x}$. The physical nature of structural effects on surface reactivity is to reveal $G_B(x)$.

**Electronic principle of size effect**

Surface activity radically originates from the residual bonding capabilities of dangling bonds of surface atoms, and can be decreased through surface reconstruction, aggregation, or adsorption[81]. These changes basically lead to two types of surface activities. One is thermodynamic activity based on atomic migrations in surface reconstruction, aggregation, sintering, and phase transition. Such activities are usually measured in terms of transition temperatures[14-16,85]. The other is chemical reactivity through adsorbing adsorbates particularly via chemisorption, and is measured by adsorption strength[72]. In addition, catalytic activity is another type of surface activity involving adsorption, activation, reaction, electron transfer, and desorption of reactants, intermediates, and products, which is intrinsically measured by the turnover frequency (TOF) of reactants[17]. Thermodynamic activity and chemical reactivity directly correlate to the bonding features of surface atoms and the electronic structures of the whole particles[79].

Size reduction can commonly enhance thermodynamic activity and chemical reactivity of nanomaterials regardless of metals or compounds. This general trend indicates there must exist a unified principle. Many size-dependent phenomena have been explained in terms of size, defect, lower coordination, and oxidation state, or studied with DFT calculations[19,20,69,73]; however, these points cannot unify the physicochemical pictures of nanomaterial surface science, nor really reveal the intrinsic roles of size. Two critical issues need to be addressed: (i) the electronic mechanism how size correlates to surface reactivity as the main variable; (ii) how size influences the effects of other structural parameters such as defect and coordination number on surface reactivity.

Both enhanced thermodynamic activity and chemical reactivity originate from reduced confinement degrees of surface atoms by the whole nanoparticles, and the electronic nature lies in weakened bonding strengths with neighboring atoms[79,85-87]. In general, weaker orbital overlaps with surrounding atoms lead to weaker confinements of surface atoms but increased redistribution freedoms for SVAOs to form surface coordination bonds. This electronic-level interaction principle can be described by $G_B(x)$. $G_B$ is the CEA measuring the constrain strength of SVAOs by surrounding atoms, thus the physical nature of size effect is to reveal how $G_B(r)$ correlates to surface activities.

Equation (6) can be expanded as

$$f_S = f_{S_0} + \Sigma \int f_S f_B \left( \frac{\dot{G}_S}{G_S} - \frac{\dot{G}_B}{G_B} \right) dx_i \quad (8),$$

where $f_{S_0}$ presents intrinsic surface reactivity on the surfaces of perfect single crystals. The expanded items express various structural effects on surface reactivity. Define

$$X = \frac{\dot{G}_S}{G_S} - \frac{\dot{G}_B}{G_B} \quad (9),$$

then

$$\Delta f_S = \Sigma \int f_S f_B \left( \frac{\dot{G}_S}{G_S} - \frac{\dot{G}_B}{G_B} \right) dx_i \quad (10).$$

The value of $X$ controls the enhancement or weakening effects of structural factors on surface reactivity. As $G_S$ and $G_B$ describe the bonding capabilities of adsorbates and nanomaterials, they are mathematically orthogonal. For given adsorbates, $\dot{G}_S=0$, then $X$ becomes

$$X = -\frac{\dot{G}_B}{G_B} \quad (11).$$

Equation (11) illustrates the intrinsic physical principle how structural parameters alter the surface reactivity of nanomaterials. Positive $\dot{G}_B$ yields negative $X$, then increase of the structural parameter leads to increased $G_B$ but decreased surface reactivity, such as size and coordination number. Conversely, negative $\dot{G}_B$ yields positive $X$, then increase of the structural parameter leads to decreased $G_B$ but increased surface reactivity, such as defects.

The surface reactivity can be expressed as

$$f_S = \frac{f_{S_0}}{f_{S_0} + \frac{G_B f_{B_0}}{G_{Bm}}} \quad (12)$$

where $G_{Bm}$ is the maximum of $G_B$. The key point to understand the electronic nature how size reduction enhances surface reactivity lies in revealing the function $G_B = G_B(r)$. For spherical nanoparticles of nonlayered materials, $G_B(r)$ follows

$$G_B = G_{Bm}(1 - \frac{kd_0}{r}) \quad (13),$$

where $d_0$ is atom size and $k$ a structure-dependent factor[79,86,87]. Then surface reactivity can be expressed as

$$f_S = \frac{f_{S_0}}{1 - \frac{kf_{B_0}d_0}{r}} \quad (14).$$

$X$ becomes

$$X = -\frac{\dot{G}_B}{G_B} = -\frac{\dot{G}_B}{G_{Bm}(1-\frac{kd_0}{r})} \quad (15).$$

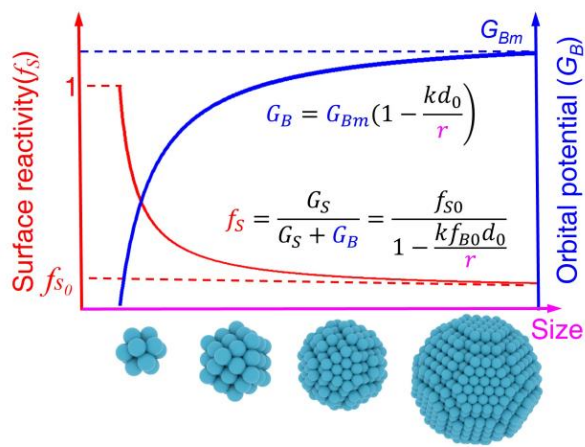

**Figure 5.** The relationships of size with nanomaterial orbital potential ($G_B$) and surface reactivity ($f_s$) derived based on competitive orbital redistribution model.

Equations (14) and (15) reveal two roles of size reduction in enhancing the surface reactivity of nanomaterials. Equation (14), derived from (6), (12), and (13), illustrates the direct role through weakening the confinement degrees of bulk energy bands to SVAOs. Decrease of $G_B$ enables more orbital overlaps in surface chemisorption bonds. This weakened lattice confinement effect on SVAOs is the primary electronic mechanism for size reduction to generally enhance the surface reactivity of nanomaterials. Equations (6), (12), (13), and (14) also reveal the electronic principle why surface reactivity is inversely proportional to particle size as shown in Figure 5. On the other hand, equation (15), derived from (9), (10), (11) and (13), illustrates the amplification effect of size reduction on modified surface reactivity caused by other parameters such as defects and coordination numbers. Defects can usually enhance surface reactivity through negative $\dot{G}_B$, but the enhancement effects are greater on nanoparticles than on single crystals, owing to the decreased $G_B$ shown by equation (13). It is difficult to experimentally probe this amplification effect owing to the simultaneously entangled impacts of all structural factors. In conclusion, weakened orbital confinement effect and amplification effect are two intrinsic electronic principles fundamentally driving size effects on enhanced surface reactivity of nanomaterials.

### Electronic principle of surface effect through nanoscale cooperative chemisorption

Surface ligands can commonly modify the properties of nanomaterials through chemisorption, regardless of their compositions and structures, and can be enhanced through decreasing particle size and increasing ligand coverage[4,5,29]. However, particle size and ligand coverage cannot in-depth reveal underlying physical principles, because the modified physical and chemical properties more basically result from perturbed electronic structures. Therefore, the key point to understand the physical nature of LISEs lies in revealing how ligands modify the electronic states of nanomaterials. To this end, these issues need to be clarified: (i) the electronic features of ligand-capped nanomaterials; (ii) how ligands modify the electronic states of nanomaterials; (iii) the roles of size, $S/V$, and ligand coverage ($\theta$) in LISEs. These issues can be unified by nanoscale cooperative chemisorption (NCC), a theoretical model we have developed to reveal the electronic model and principle of coverage-dependent ligand-nanomaterial interplays[28].

Ligands interact with nanomaterials through forming surface coordination bonds, in which SVAOs are partly redistributed from extended energy bands into localized surface bonds. Such competitive orbital redistributions of SVAOs from energy bands into surface bonds are the electronic mechanism how ligands modify the electronic structures and properties of nanomaterials. As a result, SVAOs simultaneously exist in two electronic states, energy bands and surface bonds. The two competitive electronic states are coupled with SVAOs, but are also spatially separated—this bond-band competition model is the general electronic feature of ligand-capped nanomaterials as shown in Figure 4B.

Energy bands, formed through extended orbital overlaps within the lattices, are the primary electronic features of nanomaterials. Density of states (DOSs) is a critical concept describing the features of energy bands. In general, more VAOs and higher overlap degrees lead to wider DOSs composing of more continuous energy levels, while less VAOs and lower overlap degrees yield narrower DOSs with discrete energy levels. Stronger energy bands can more strongly confine SVAOs, yielding greater $G_B$. LISEs on large nanoparticles and single crystals are usually neglectable, because ligands cannot effectively alter their energy bands. For nanomaterials smaller than certain critical sizes, however, their properties can be highly affected or even dominated by surface ligands, because their energy bands can be more readily tuned owing to limited orbital numbers.

Figure 6A illustrates coverage-dependent distribution states of SVAOs[28]. For nanomaterials without surface ligands (i), SVAOs preferentially overlap with the VAOs of lattice atoms. This yields wider energy bands as shown by the DOS in Figure 6B, which are the intrinsic electronic structures of nanomaterials. When an adsorbate adsorbs on a surface site, its SVAO will be polarized from energy bands to form a surface bond (ii). The result is to redistribute the SVAO from energy bands into the surface bond. This redistribution decreases the overlap degrees of orbitals in energy bands. Increasing ligand coverage will redistribute more SVAOs into surface bonds, which further decreases overlapped orbitals in energy bands (iii). At adsorption saturation, all surface sites are capped by ligands (iv). In this case, SVAOs are maximally redistributed into surface bonds, yielding the weakest orbital overlaps in energy bands. This leads to the narrowest energy bands and maximally quantized band levels (Figure 6B).

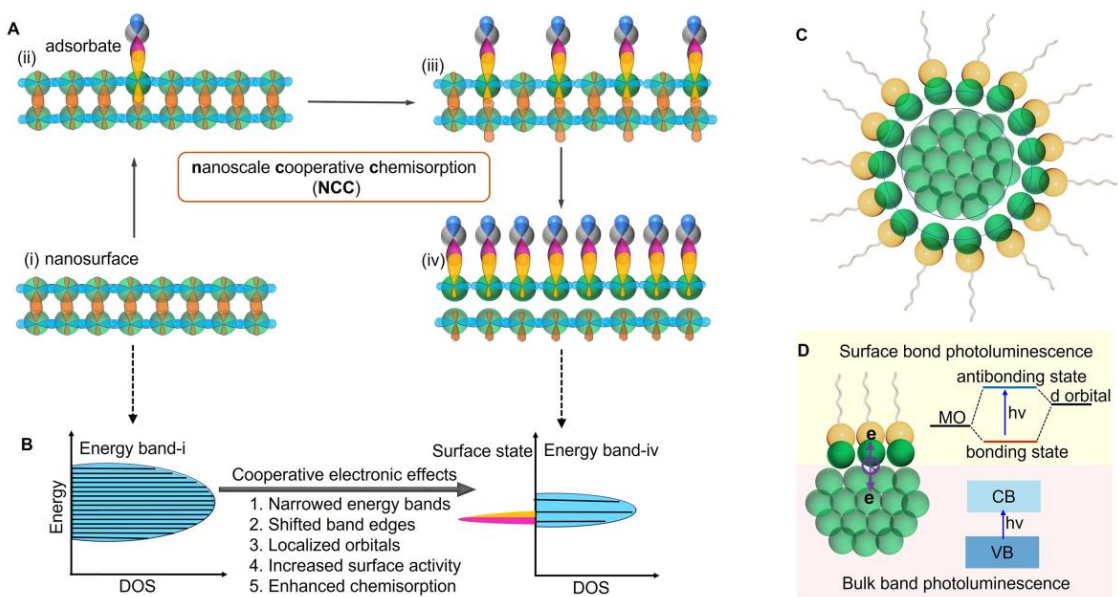

**Figure 6.** The model of coverage-dependent nanoscale cooperative chemisorption (NCC). (**A**) Enhanced orbital redistributions from lattices into surface bonds with increased adsorbate coverages. (**B**) Modification of nanomaterial DOS electronic states by surface ligands from clean surface to saturation adsorption. (**C**) Saturation chemisorption of ligands leads to weakly-bonded surface layers with inner cores of nanoparticles, in which the bonding states of surface atoms are different from inner atoms. (**D**) Blocked exciton diffusions between surface layers and inner cores of ligand-capped nanoparticles. This leads to enhanced photoluminescent quantum yields resulting from both surface bonds and bulk bands.

The effects of surface ligation on the electronic structures of nanomaterials are to narrow DOSs and quantize band levels. In this process, the narrowed bands and quantized band levels by ligation can further decrease the confinement degree of SVAOs by energy bands, which further increases surface reactivity in turn. As a result, ligands will be more spontaneously adsorbed on nanomaterials, which leads to even narrowed DOSs and further quantized band levels. This is a positive feedback effect and will spontaneously lead to saturation adsorption. Therefore, the trend is that increasing ligand coverage can incrementally enhance chemisorption until saturation. This coverage-enhanced ligand-nanomaterial interaction trend is referred as nanoscale cooperative chemisorption (NCC), in which the cooperative effect refers to enhanced chemisorption strength with increased ligand coverages[28].

Narrowed energy bands and even quantized band levels are the intrinsic electronic principles how surface ligands modify the physical and chemical properties of nanomaterials. The narrowed energy bands directly shift band edges such as valence band maximum (VBM) and conduction band minimum (CBM) of semiconductor nanostructures, which further increases band gaps and the energies of optical absorption and emission. Other physical properties will also be modified owing to weakened orbital couplings. The even quantized band levels can further weaken orbital confinements, leading to decreased $G_B$ and increased surface reactivity. In particularly, the strength of chemisorption is the strongest at saturation states. This electronic mechanism elucidates the physical nature why nanocrystals tend to be saturated by surface ligands[62,88], and why the adsorption of certain ligands can modify the catalytic activity and selectivity of nanocatalysts[89-91].

### Unified electronic principle of nanomaterial optical properties

NCC model concludes that saturation adsorption yields the strongest chemisorption interactions between ligands and nanomaterials, owing to the maximized confinement of SVAOs in localized surface bonds. As a result, surface atoms display minimized orbital overlaps with other neighboring lattice atoms (Figure 6A iv). This orbital-level polarization effect separates surface atoms from inner bulk atoms, forming a type of core@shell structure (Figure 6C). The core consists of stoichiometric inner atoms of nanomaterials, and displays bulk energy bands. The shell consists of complexes of ligands with surface atoms, and forms localized surface bond states. This core@shell polarization effect can be enhanced with increased chemisorption strengths. The model can uniformly interpret size- and coverage-dependent optical properties of both surface and bulk fluorescence[67,92-94].

The photoluminescence of Au NCs originates from localized surface bonds, thus the emission energy is determined by the energy difference between bonding and antibonding states (Figure 6D). Increasing ligand coverage can enhance chemisorption, meaning that orbital overlaps between Au and S atoms are increased. According to the principles of molecular orbital theory, increased orbital overlaps lead to increased energy gaps between bonding and antibonding states. Therefore, increasing ligand coverage can increase the emission energy of Au NCs. This mechanism accounts for why increasing GSH coverage can shift the photoluminescent wavelengths of Au NPs from 810 to 600 nm[42].

Weakened orbital overlaps between inner cores and surface shells account for the coverage-dependent photoluminescent quantum yields of both surface bond and bulk band fluorescence. Excitons can diffuse between surface atoms and bulk atoms. When bulk excitons diffuse into surface atoms like in QDs, they will be trapped by the dangling orbitals of surface atoms, leading to decreased PLQYs. Reversely, when surface excitons diffuse into bulk phase like in ligand-capped Au NCs, they will dissipate as heat, also leading to decreased PLQYs[49,95]. The weakened orbital overlaps of SVAOs with inner core atoms at high ligand coverages can block the diffusion pathways of electrons between the inner cores and shells (Figure 6B). As a result, the excited electrons will be confined in surface bonds or inner cores. This polarization effect fundamentally enhances PLQYs of both surface fluorescence such as Au NCs and bulk fluorescence such as QDs and rare earth nanoparticles[34,42,50-52].

It has been widely found that PLQYs of QDs can be tuned through engineering their surface chemistry[42,50,51]. Metal chalcogenide nanocrystals can be thought as a stoichiometric core with a layer of metal-ligand complexes adsorbed to their surfaces, and both the core stoichiometry and ligand coverages can affect PLQY. Successive ionic layer adsorption and reaction studies have shown that cadmium chalcogenide (CdE, X=S, Se, and Te) nanocrystals with cadmium-rich surface layers are generally brighter than chalcogenide-rich nanocrystals[25,44]. These phenomena were interpreted by the roles of surface metal ions to prevent charge trapping. However, the dangling orbitals of surface metal atoms are actually the trapping states for excited electrons. The real mechanism is that more metal atoms on the surfaces can increase ligand coverages to maximize the effects of nanoscale competitive chemisorption, and then generate even weaker orbital couplings between surface complex layers and inner cores.

Nanoscale cooperative chemisorption (NCC) reveals the general electronic principle how ligands interact with nanomaterials. This model is based on the competition between extended energy bands and localized surface bonds in occupying SVAOs. The theory can unify the physical nature why reducing particle size and increasing ligand coverage can enhance chemisorption strength and how ligands modify the electronic structures and properties of nanomaterials.

### On the roles of size and ligand coverage

NCC model reveals the mechanisms how ligands modify the electronic structures of nanomaterials through competitive orbital redistribution. The roles of size and $S/V$ of nanomaterials and capping atom and coverage of ligands ($\theta$) lie in how they affect the modification effects of nanomaterial electronic structures. In surface chemical interactions, the direct electronic attribute determining surface reactivity and LISEs is the confinement strength of SVAOs by lattice atoms and energy bands, that is $G_B(x)$. In the competition processes of surface orbital redistributions, surface ligands are cooperative in weakening orbital couplings in energy bands as revealed by NCC model. On the other hand, orbital couplings in energy bands can be enhanced by increasing lattice atoms (that is particle size), which can be approximatively illustrated by equation (13). The competition capabilities of ligands and nanomaterials in attracting SVAOs in surface bonds positively correlate with the numbers of capping atoms ($m$) and lattice atoms ($n$), following

$$\frac{m}{n} = \frac{\theta S}{V} \quad (16).$$

Chemisorption strength through orbital overlaps is fundamentally determined by the attributes of overlapped atomic orbitals. The formation of localized surface bonds follows the conditions of molecular orbitals. Combining the results of molecular orbital theory and the features of nanoscale cooperative chemisorption, coverage-dependent chemisorption strength follows

$$f_S \propto \frac{\theta S}{V}\left(1 - \frac{\Delta}{\sqrt{\Delta^2 + 4\beta^2}}\right) \quad (17),$$

where $\Delta$ is energy level difference between SVAO and ligand's HOMO, and $\beta$ is resonance integral measuring the strength of bonding interactions as a result of orbital overlaps.

Ligand is the driving force to change the electronic structures of nanomaterials, while the bulk atoms apply resistance against such changes. Equation (17) approximately shows the interaction trend of surface atom, particle size, $S/V$, capping atom, and ligand coverage in enhancing surface chemical interactions on nanomaterials. Size ($V$) measures the confinement strength of SVAOs by nanoparticle energy bands; $S$ measures the maximum degree of ligands to interact with nanomaterials. Ligand coverage ($\theta$) shows dynamic and cooperative features, while the product of $\theta S$ denotes ligand number. It is noted that surface chemical interactions are intrinsically controlled by the attributes of SVAOs and ligands' frontier molecular orbitals, which are illustrated by $\Delta$ and $\beta$. In summary, $\Delta$ and $\beta$ are the intrinsic electronic attributes determining the formation and strength of surface chemisorption interactions, while $S/V$ and $\theta$ are extrinsic factors further modifying the degrees of orbital overlaps. It is noted that equation (17) is not strictly deduced based on NCC model. More fundamental insights still need to be explored to understand how $G_B$ depends on $S/V$ and $\theta$.

### Experimental probing of competitive orbital redistribution

The direct electronic results of adsorbate-nanomaterial interactions are to narrow and quantize energy bands through competitively redistributing SVAOs from energy bands into surface bonds. In practice, however, it is highly challenging to experimentally probe such changed electronic states. This is limited not only by ideal ligand-nanomaterial chemisorption models, but also by applicable characterization methods. For example, although the photoluminescent performances of ligand-capped Au NCs and QDs can be tuned by varying particle sizes and ligand coverages, it is difficult to probe the changes in their bulk energy bands with X-ray photoelectron spectroscopy (XPS) and X-ray absorption fine structure (XAFS). Such experimental challenges have severely limited in-depth insights into the electronic principles underlying nanomaterial surface science.

To effectively identify the varied electronic states caused by surface ligands, the interferential signals from the bulk atoms should be suppressed, because chemisorption can only perturb the electronic states of top surface atoms. Atomically thin two-dimensional (2D) nanosheets are ideal model systems, because their extremely large specific surface areas can maximize surface ligation sites and the modification of energy bands. We have shown TiO$_2$ nanosheets are suitable systems to experimentally explore the electronic mechanisms of nanomaterial surface science, because the varied conduction bands mainly composed of 3d orbitals can be revealed by probing the changes of Ti-L edges with near-edge X-ray absorption fine structure (NEXAFS). Figure 7 presents the coverage-dependent optical absorption and electronic states of 0.37-nm-thick TiO$_2$ nanosheets modified by ethylene glycol (EG)[28].

TiO$_2$ nanosheets were prepared by hydrolyzing TiCl$_4$ in EG, in which EG acts as bidentate capping ligands to stabilize the (010) facet (Figure 7A). The thickness is one unit in the $b$ direction (~0.37 nm), containing 3 Ti-O layers (Figure 7B). Through varying EG coverages, the band gap ($E_g$) can be tuned between 3.24 and 3.62 eV. For nanosheets saturated by EG, $E_g$ is 3.62 eV, but the value decreases to 3.24 eV after removing 63% EG ligands, which is the same to the bulk $E_g$ of TiO$_2$ nanowires (Figure 7C). Band gaps of the nanosheets can be continuously tuned between 3.24 eV and 3.62 eV through controlling EG coverages (Figure 7D). $E_g$ of the TiO$_2$ nanosheets increases with increased EG coverages, clearly displaying coverage-dependent optical absorption. Such surface effects become more prominent when $\theta > 0.90$. These results show that ligands cannot effectively modify the electronic structures and properties of nanomaterials at low coverages, but can significantly dominate the properties at high coverages[28].

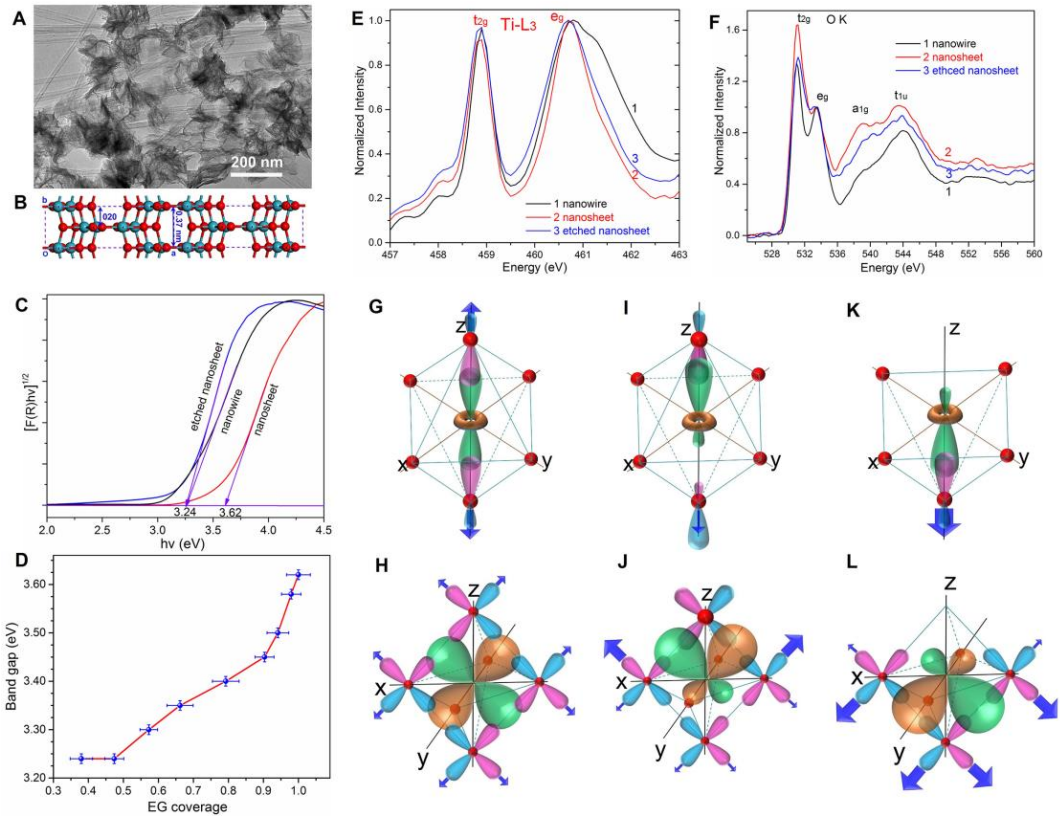

**Fig. 7.** Experimental example of coverage effect on the optical absorption and electronic structure of TiO$_2$ nanosheets. (**A**) TEM image of TiO$_2$ nanosheets prepared by hydrolyzing TiCl$_4$ in ethylene glycol (EG). (**B**) Atomic structure model of TiO$_2$ nanosheets exposing (010) facet. (**C**) Optical absorption spectrum of TiO$_2$ nanomaterials. (**D**) The dependence of band gaps on EG coverage. NEXAFS results of Ti-L$_3$ (**E**) and O-K (**F**) edges. Distribution configurations of surface Ti 3$d$ orbitals in octahedral ligand fields with different ligands in (**G, H**) nanowires, (**I, J**) nanosheet, and (**K, L**) etched nanosheets. (**G, I, K**) show varied bonding features of $\sigma$-type 3$dz^2$; (**H, J, L**) show varied bonding features of $\pi$-type 3$d_{xz}$ and 3$d_{yz}$.

The modifications of energy bands can be revealed by probing Ti-L$_3$ and O-K edges with NEXAFS[96]. Figures 7E and 7F show Ti-L$_3$ and O-K lines of (1) TiO$_2$ nanowire, (2) TiO$_2$ nanosheet saturated by EG, and (3) etched TiO$_2$ nanosheet with EG coverage of 0.40. The varied width of $e_g$ and intensity of $t_{2g}$ peaks reflect different band states. Compared to nanowire and etched nanosheet, the narrowed Ti-L$_3$ $e_g$ peak of EG-saturated nanosheets suggests decreased overlap degrees of $d_{x^2-y^2}$ and $d_{z^2}$ orbitals in the lattices (Figures 7E). This is because the d orbitals are redistributed from energy bands into surface bonds with EG as shown by Figure 7I and 7J. At low EG coverages, the 3$d$ orbitals preferentially extend into the lattice to form energy band states (Figure 7K and 7L), which leads to wider $e_g$ and stronger $t_{2g}$ bands. Meanwhile, EG ligands can also form localized Ti-O bonding states as shown by the stronger $t_{2g}$ peak in O-K line. These results demonstrate that surface ligands can modify the electronic structures of nanomaterials through competitively redistributing SVAOs, and provide experimental supports for NCC model.

## Conclusion and outlook

Many disciplines meet in the surface science of nanomaterials owing to the wide foundation roles in adsorption, catalysis, nanocrystal synthesis and stabilization, and optical absorption and emission. Understanding the intrinsic physical and chemical principles how particle size, ligand, and ligand coverage generally affect the properties of nanomaterials has been long remaining challenging. We have clarified how size and ligands modify the energy band states of nanomaterials, and how the attributes of energy bands determine surface reactivity. Surface ligands interact with nanomaterials through competitively redistributing surface atomic orbitals from energy bands into surface bonds. Nanomaterials therefore spatially display two competitive states, energy bands extended in the bulk lattices and surface bonds locally confined between adsorbates and surface atoms. Such competitive orbital distributions intrinsically drive size effect and surface effect in nanomaterial surface science.

The confinement degree of surface atomic orbitals by lattice band states, measured by $G_B$, directly determines surface reactivity. Both size reduction and surface ligation can reduce orbital overlaps in energy bands, leading to weakened band states (indicated by narrowed bandwidth) and quantized energy levels. Such modified band states intrinsically lead to decreased $G_B$ but increased surface reactivity. This trend positively correlates with $θS/V$, that is increase of both ligand coverage and $S/V$ can enhance surface reactivity. This interaction trend is general for all inorganic nanomaterials, regardless of their chemical nature of metals, oxides, semiconductors, or layered compounds, while the difference lies in the specific expressions of $G_B(x)$. Exploring the expressions of $G_B(x)$ functions for specific nanomaterials is the intrinsic but also challenging issue to reveal structure-property relationships in nanomaterial surface science.